\documentclass{iopart}   
\usepackage{iopams}  
\usepackage{epsf}  
\usepackage{times}  
 
\eqnobysec

\begin{document}

\jl{1}

\letter{Solvable Quantum Two-body Problem: Entanglement}

\author{ML Glasser\dag\ddag\ and LM Nieto\dag}

\address{\dag\ Departamento de F\'isica
Te\'orica, At\'omica y \'Optica,  Universidad
de Valladolid\\
47071 Valladolid, Spain}

\address{\ddag\ Center for Quantum Device Technology, Clarkson University, Potsdam\\ NY 13699-5820, USA}

\begin{abstract}
A simple one dimensional model is introduced describing a two
particle ``atom" approaching a point at which the interaction
between the particles is lost. The wave function is obtained
analytically and analyzed to display the entangled nature of the
subsequent state.
\end{abstract}


\submitted

\section{Introduction}

The notion of  entanglement was introduced by Schr\"odinger \cite{schr} who considered as the essential feature of quantum mechanics the fact that when two particles interact by a known force and then separate, they can no longer be considered as independent.  This ought to be evident in the case of a hydrogenic atom for which, by some mechanism, the interparticle interaction is lost and the atom is ionized. The entangled nature of the state must be evident in the structure of the two-particle wave function, but this does not appear to have been examined in detail. The aim of this note is to study this situation for an exactly solvable two-particle system.

 A suitable one dimensional model was introduced in \cite{glass} to describe the
interaction of a hydrogenic atom with a ``metal" surface, such
that once the atom penetrates the metal the nuclear ``charge" is
screened to zero. It was found that the problem could be re-expressed
as the Wiener-Hopf problem introduced in 1947 to describe
the reflection of an electromagnetic wave from a linear coastline  solved in 1952 by  Bazer and Karp. Their  work is
reviewed in \cite{baz} and from the results,  a formal expression for the two-body
wave function  was obtained along with an exact formula
for the reflection coefficient  as a function of
the incident energy $E$. 

In the atomic center of mass system
(total mass $M$, center of mass coordinate $R$, reduced mass $\mu$
and relative coordinate $r$) the interaction potential for the model is
$V(R,r)=-\lambda \delta(r)$ outside the metal ($R>0$)  and 0
inside ($R<0$). In terms of reduced variables:
\begin{equation}
y=\sqrt{\mu/M}r, \ \ 
k_0^2=2ME/\hbar^2,\ \ 
a=\sqrt{M\mu}\lambda/\hbar^2,\ \ 
K^2=k_0^2+a^2
\label{11}
\end{equation}
the wave function is (the sign in the exponent of the Fourier transform has been
altered from that in \cite{glass} to reflect standard usage)
\begin{eqnarray}
\fl
\psi(R,y) & = &  \exp[-\rmi KR-a|y|]+\frac{a^2[\sigma^+(K)]^2}{2K^2}\exp[\rmi KR-a|y|] \nonumber \\
\fl
&& -\frac{a\sigma^+(K)}{2\pi
\rmi }\oint\frac{\exp[-\rmi kR-|y|\sqrt{k^2-k_0^2}]}{(k+K)[a-\sqrt{k^2-k_0^2}]}\  \sigma^+(k)\  \rmd k\quad (R>0),
\label{12}
\\
\fl
\psi(R,y) & = &  \frac{a\sigma^+(K)}{2\pi
\rmi }\oint \frac{\exp[-\rmi kR-|y|\sqrt{k^2-k_0^2}]}{(k+K)\sqrt{k^2-k_0^2}}\ \sigma^-(k)\ \rmd k\quad (R<0),
\label{13}
\end{eqnarray}
where, $\sigma^{\pm}(k)$ are the Wiener-Hopf
factors \cite{Karp} of the kernel 
\begin{equation}
\sigma(k)=1-\frac{a}{\sqrt{k^2-k_0^2}}=\frac{\sigma^+(k)}{\sigma^-(k)}\label{14}
\end{equation}
and the
contours for the integrals in (\ref{12}) and (\ref{13}) surround  branch cuts
associated with $\pm k_0$ (which, for technical reasons, have  infinitesimal imaginary
parts) in the upper and lower half planes.

The leading term on the right hand side of (\ref{12}) represents an
incoming atom and is normalized to unit amplitude; the second term
represents the atom reflected from the surface with reflection
coefficient  \cite{glass}
\begin{equation}
{\cal{R}}=\frac{(K-k_0)^2}{K^2}.\label{15}
\end{equation}
The task here is to analyze the integrals in (\ref{12}) and
(\ref{13})  which describe the entanglement of the atomic particles in the
``ionized'' state. Since for $R<0$ there is no direct interaction between
the two particles comprising the atom, a dependence of
the wave function on the relative coordinate $y$ indicates the quantum
entanglement of the two particles.

 In the following section we change  (\ref{13}) and (\ref{14})  into somewhat more convenient forms and  
 express all
the components of the Wiener-Hopf solution  in terms of
standard functions. In the concluding section details of the wave
functions are presented and we find that indeed, even for $R$ large and
negative,  the two-particle wave function is not separable.

\section{Calculation}

We begin by investigating the Wiener-Hopf factorization of
$\sigma(k)$ (\ref{14}). Only the magnitude of the $+$ factor at the value
$k=K$ was needed in \cite{glass}; we now require its complex value for
${\rm Im\;} k\ge0$. The calculation is  simplified somewhat  by introducing
the modified kernel
\begin{equation}
S(k)=\frac{k^2-k_0^2}{k^2-K^2}\ \sigma(k)=\frac{S_+(k)}{S_-(k)}\label{21}
\end{equation}
in terms of which (\ref{12}) and (\ref{13})  become
\begin{eqnarray}
\fl
\psi(R,y)=\exp[-\rmi KR-a|y|]+2\left(\frac{K-k_0}{K+k_0}\right)S^2_+(K)\exp[\rmi KR-a|y|]-\Phi(R,y)\label{22}
\\
\fl
\Phi(R,y)=\frac{2aK}{K+k_0}S_+(K)\oint\frac{\rmd k}{2\pi
\rmi }\ \sqrt{\frac{k+k_0}{k-k_0}}\ \frac{\exp[-\rmi kR-|y|\sqrt{k^2-k_0^2}]}{S_+(k)(K^2-k^2)}\mbox{\hskip
.2in}(R>0),\label{23}
\end{eqnarray}
where the contour encloses the branch cut running from $-k_0$ in the lower half plane, parallel
to the real axis to the  imaginary axis and then to $-i\infty$.
For $R<0$,
\begin{equation}
\fl
\psi(R,y)=\frac{2aKS_+(K)}{K+k_0}\oint\frac{\rmd k}{2\pi
\rmi }\  \sqrt{\frac{k+k_0}{k-k_0}}\
\frac{\exp[-\rmi kR-|y|\sqrt{k^2-k_0^2}]}{S_+(k)(K^2-k^2)}\label{24}
\end{equation}
where the contour surrounds the branch cut running from $k_0$, parallel to the real axis up to the imaginary axis and
then to $i\infty$. An alternative derivation of (\ref{22})--(\ref{24}) is outlined in Appendix A.

Now \cite{baz,nob},
\begin{eqnarray}
S_+(k)&=&\exp[-J(k)]  \nonumber\\
J(k)&=&\frac{1}{2\pi
\rmi }\int_{-\infty}^{\infty} {\rm Log}\left[1+\frac{a}{\sqrt{u^2-k_0^2}}\right]\frac{\rmd u}{u-k},\label{25}
\end{eqnarray}
where the contour runs along the real axis indented below $u=k$.
By breaking the range into $ [-\infty, 0]\cup [0,\infty]$ and
combining the two integrals we have
\begin{equation}
J(k)=\frac{k}{\pi
\rmi }\int_0^{\infty}{\rm Log}\left[ 1+\frac{a}{\sqrt{u^2-k_0^2}}\right]\ \frac{\rmd u}{u^2-k^2}.
\label{26}
\end{equation}
Next, by letting $u\rightarrow -\rmi s$ and, since there are now no
singularities in the first quadrant, rotating the contour back to the
real axis,  with $s\rightarrow ks$, we find
\begin{eqnarray}
\fl
J(k)&=&\frac{1}{\pi}\int_0^{\infty}\frac{\rmd s}{s^2+1}{\rm Log} \left[ 1-\frac{\rmi a}{k\sqrt{s^2+(k_0/k)^2}} \right] \nonumber\\
\fl 
&=& \frac{1}{\pi}\int_0^{\infty}\frac{\rmd s}{s^2+1}{\rm Log}[\sqrt{s^2+(k_0/k)^2}-\rmi a/k]- \frac{1}{2\pi}\int_0^{\infty}\frac{{\rm Log}[s^2+(k_0/k)^2]}{s^2+1}\ \rmd s.
\label{27}\end{eqnarray}
The second integral, as can be found in tables, is $\pi
{\rm Log}[1+(k_0/k)]$; the first integral is evaluated in  Appendix B yielding
\begin{equation}
S_+(k)=\sqrt{\frac{k+k_0}{k+K}}\ \exp\left[\frac{2\rmi }{\pi}\ {\rm Im\; Ti}_2 \left( \frac{\sqrt{k_0^2-k^2}+\rmi a}{K+k}\right) \right].
\label{28}
\end{equation}
where ${\rm Ti}_2(z)$ is the Arctangent integral function, defined in (\ref{ti2}).
Formula (\ref{28}) is valid for ${\rm Im\;} k\ge0$, except for $k=\pm K$, where
there is a confluence of singularities. However, this case can be
extracted from the results given in \cite{glass} whence we find
\begin{eqnarray}
S_+(K)=\sqrt{\frac{K+k_0}{2K}}\ \exp\left[ \frac{\rmi}{2\pi}\,   \biggl({\rm Li}_2(-a/K)-{\rm Li}_2(a/K) \biggr)\right].
\label{29}\\
S_+(K)S_+(-K) =\frac12.
\end{eqnarray}

\section{Results and conclusions}

We turn now to the wave function (\ref{24}) in the interaction-free region. At this point the imaginary part of  $k_0$ can be set  to zero and on the part of the
contour surrounding the vertical portion of the branch cut, where the factor
$\exp(-\rmi kR)$ is a decaying exponential, the contribution to the integral will be small and we shall neglect it. Similarly, since the contribution of the contour in the upper half plane about the interval $[0,k_0]$ is exponentially small compared to the part in the lower
half plane, for $R<0$ we have
\begin{equation}
\fl
\psi(R,y)\approx\frac{2aK}{K+k_0}S_+(K)\int_0^{k_0}\frac{\rmd k}{2\pi}\ \sqrt{\frac{k_0+k}{k_0-k}}\
\frac{\exp[-\rmi kR-\rmi |y|\sqrt{k_0^2-k^2}]}{S_+(k)(K^2-k^2)}.\label{31}
\end{equation}

Note that, from (\ref{15}), as $k_0\to 0$,
${\cal{R}}=1$ signifying total reflection. Accordingly, from (\ref{31}),
$|\psi(R,y)|^2=0$ apart from a rapidly decaying evanescent wave
due to the neglected part of the contour.

\begin{figure}
\begin{center}
\epsfbox{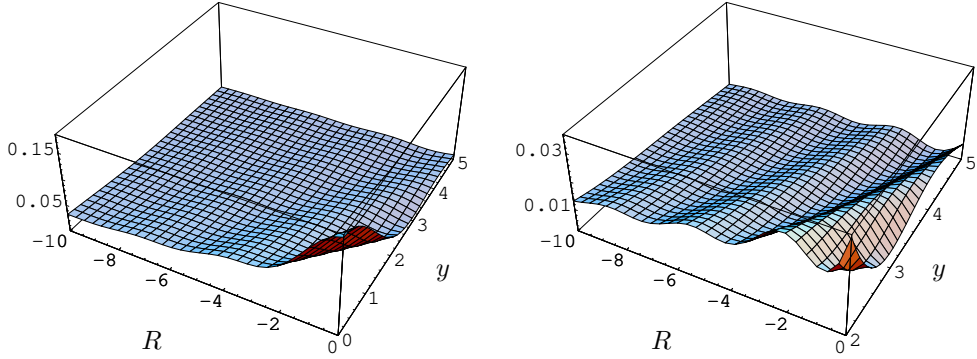}
\end{center}
\caption{Square of the wave function amplitude $|\psi(R,y)|^2$: $a=1$, $k_0=2$}
\end{figure}

We have evaluated (\ref{31}) numerically for $a=1$, $k_0=2$, and present the square of its amplitude as a function of $|y|$ and $|R|$ for $a=1$ and $k_0=2 $ in Fig.~1. In Fig.~2 we show the real and imaginary parts of $\psi(R,y)$.
For $R$ large and negative, but $| y|\ll | R|$, the integral can be estimated by
setting $k=0$ in the integrand, apart from the factor
$\exp(-ikR)$, leading to
\begin{eqnarray}
\psi(R,y)\approx\frac{a}{\rmi \pi K^2R\sqrt{2k_0(K+k_0)}}\ e^{-\rmi [k_0|y|+\phi_-]}\label{32} \\
\phi_- =\frac{2}{\pi}\ {\rm Im\; Ti}_2 \left( \frac{k_0+\rmi a}{K}\right).
\nonumber
\end{eqnarray}
Thus, the amplitude decays as $| R|^{-1}$ and even far away from the point where the particles are decoupled, the relative coordinate is present in the phase.  {
 For larger values of $|y|$, the amplitude appears to be ${\rm o}(|y|^{-3/4})$ and oscillates as $y\to\infty$, which is illustrated in Fig.~3 showing $|\psi(-10,y)|^2$ for $a=1$, $k_0=2$.




A natural quantity to examine  is the expected value
\begin{equation}
Y(R)=\frac{\int_{-\infty}^{\infty}|y||\psi(R,y)|^2 \rmd y}{\int_{-\infty}^{\infty} |\psi(R,y)|^2 \rmd y}
\label{34}
\end{equation} 
of the relative particle displacement . Our numerical integration of the numerator in( \ref{34}) appears to diverge in the interaction-free region $R<0$.  To obtain a reliable result will require a more detailed asymptotic 
study of the wave function for $|y|/R\rightarrow\infty$.
Finally, in Figure~4 we show the square amplitude of the wave function (\ref{12}) for $R>0$ with $a=1$ and $k_0=2$ (including the
incident wave).

\begin{figure}
\begin{center}
\epsfbox{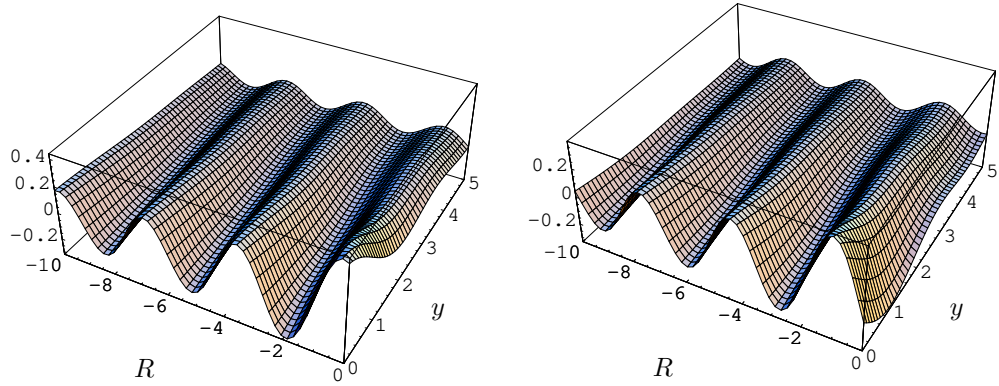}
\end{center}
\caption{Real (left)  and Imaginary (right) parts of $\psi(R,y)$: $a=1$, $k_0=2$}
\end{figure}

For $R\gg 1$ the integral in (\ref{13}) has been treated by Kay \cite{baz} who
used the method of steepest descent. From his results we find for
the integral in (\ref{23}),
\begin{eqnarray}
\fl 
\Phi(R,y)\approx\frac{\rmi^{3/2}k_0a}{4\pi^{1/2}(k_0R)^{1/2}}
\frac{\xi^2}{a^2+K^2\xi^2}\sqrt{\frac{k_0}{K}\frac{K+k_0}{K+\frac{1}{2}k_0\xi^2}}
\  e^{\rmi [k_0R(1-\frac{1}{2}\xi^2)+\frac{2}{\pi}\phi_+]}\label{35}
\\
\fl 
\phi_+=\frac{1}{2}\, {\rm Li}_2(a^2/K^2)-2 {\rm Li}_2(a/K)-{\rm Im\; Ti}_2 \left(\frac{k_0\xi+\rmi a}{K-k_0+
\frac{1}{2}k_0\xi^2}\right) \nonumber
\end{eqnarray}
where $\xi=y/R<1$. This represents an
outgoing wave with no trace of the factor $\exp(-a|y|)$
representing the atomic state.

\begin{figure}
\begin{center}
\epsfbox{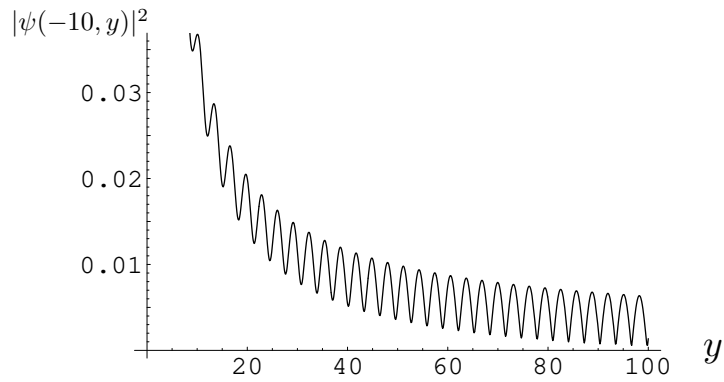}
\end{center}
\caption{$|\psi(-10,y)|^2$: $a=1$, $k_0=2$}
\end{figure}


\begin{figure}
\begin{center}
\epsfbox{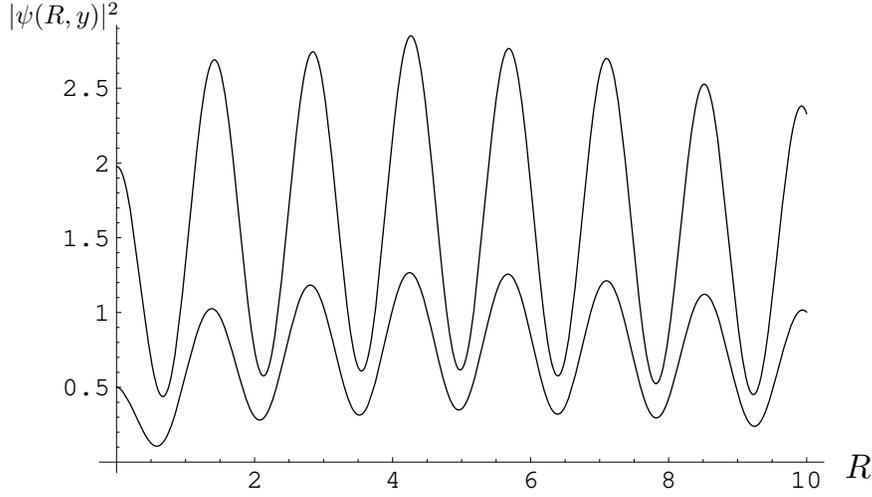}
\end{center}
\caption{$|\psi(R,0)|^2$ (upper curve) and $|\psi(R,.5)|^2$ for $R>0$: $a=1$, $k_0=2$}
\end{figure}

\ack 

This work is supported by the Spanish MEC (BFM2002-03773 and MLG grant SAB2003-0117) and Junta de Castilla y Le¥on (VA085/02). MLG thanks the Universidad de Valladolid for hospitality and the NSF (USA) for partial support (DMR-0121146).

\vskip1cm

\newpage

\vskip1cm

\appendix
\section{}

   The two-particle  Schr\"odinger equation is
\begin{equation}
L\psi(R,y)=V(R,y)\psi(R,y)
\label{a1}
\end{equation}
   where $L=\partial_R^2+\partial_y^2+k_0^2$ and has the Green function $G(R-R',y-y';k_0)$, which in momentum space is simply $[P_R^2+p_y^2-k_0^2]^{-1}$

   The solution of (\ref{a1}) is
\begin{equation*}
\psi(R,y)=\int_{-\infty}^{\infty}\int_{-\infty}^{\infty}G(R-R'y-y';k_0)V(R',y')\psi(R',y')\ \rmd R'\, \rmd y'.
\end{equation*}
   In terms of $h(R)=\psi(R,0)$ this reduces to
\begin{equation}
\label{a2}
\psi(R,y)=2a\int_0^{\infty}G(R-R',y;k_0)h(R')\, \rmd R'
   \end{equation}
   and, in particular,
\begin{equation}
\label{a3}
h_+(R)+h_-(R)=2a\int_{-\infty}^{\infty}G(R-R',0;k_0)h_+(R')\, \rmd R'
\end{equation}
   where $h_+(R)=h(R)\Theta(R)$, $h_-(R)=h(R)\Theta(-R)$. The Fourier transforms of these functions $H_+(k)$ and $H_-(k)$ are analytic in the upper and lower half $k$-plane, respectively. Since the integral in (\ref{a3}) is a convolution, by taking the Fourier transform  (\ref{a3}) becomes
   \begin{equation}
H_+(k)+H_-(k)=2ag_0(k)H_+(k)
\label{a4}
\end{equation} 
where $g_0(k)=1/2\sqrt{k^2-k_0^2}$ is the Fourier transform of $G(R,0;k_0)$. Therefore,
   \begin{equation}
\label{a5}
\frac{H_-(k)}{H_+(k)}=\frac{k^2-K^2}{k^2-k_0^2}S(k)
\end{equation}
   where
   \begin{equation}
\label{a6}
S(k)=\frac{k^2-k_0^2-a\sqrt{k^2-k_0^2}}{k^2-K^2}=\frac{S_+(k)}{S_-(k)}
\end{equation}
   and $S_+(k) (S_-(k))$ is analytic and free of zeros for ${\rm Im\;} k>0$ (${\rm Im\;} k<0$). Equation (\ref{a6}) implies that $(k^2-K^2)S_+(k)H_+(k)/(k+k_0)$ is a bounded entire function and is therefore a constant $\alpha$.
   Consequently,
   \begin{equation}
\label{a7}
h_+(R)=\frac{\alpha}{2\pi}\int_{-\infty+\rmi c}^{\infty+\rmi c}\frac{k+k_0}{S_+(k)(k^2-K^2)}\ e^{-\rmi kR}\ \rmd k
\end{equation}
   with ${\rm Im\;} K<c<{\rm Im\;} k_0$. Thus, we have from (\ref{a2}), for $-\infty<R<\infty$,
   \begin{equation}
\label{a7}
\psi(R,y)=\frac{a\alpha}{2\pi \rmi }\int_{-\infty+\rmi c}^{\infty+\rmi c} \rmd k\ \sqrt{\frac{k_0+k}{k_0-k}}\
   \frac{e^{-\rmi kR-|y|\sqrt{k^2-k_0^2}}}{(k^2-K^2)S_+(k)}.
\end{equation}
   For $R>0$ the contour can be closed into the lower half plane, which contains the two poles $\pm K$ and the branch point $-k_0$ yielding (\ref{22})--(\ref{23}), while for $R<0$ the contour can be closed into the upper half plane yielding (\ref{24}).

\section{}

Here we evaluate the integral
   \begin{equation}
\int_0^{\infty}\frac{{\rm Log}[\sqrt{x^2+c^2}+d]}{x^2+1}\ \rmd x
   \end{equation}
needed to obtain $S_+(k)$. We begin indirectly by setting
   \begin{equation}
A=\frac{y\cot(\theta/2)}{1+\sqrt{y^2+1}}\mbox{ \hskip .2in}
B=\frac{y\tan(\theta/2)}{1+\sqrt{1+y^2}}.
   \end{equation}
 Then
$$\frac{A+B}{1-AB}=y\csc(\theta)$$
and
$$\tan^{-1}(y\csc(\theta))=\tan^{-1}(A)+\tan^{-1}(B).
$$
But,
\begin{eqnarray*}
\csc(\theta)\tan^{-1}(y\csc(\theta))=-\frac{d}{d\theta}\int_1^{\csc(\theta)}\frac{\tan^{-1}(yu)}{\sqrt{u^2-1}}\ \rmd u \\
\csc(\theta)\tan^{-1}(A)= -\frac{d}{d\theta}\int_0^A\frac{\tan^{-1}(u)}{u}\ \rmd u  \\
\csc(\theta)\tan^{-1}(B)= \frac{d}{d\theta}\int_0^B\frac{\tan^{-1}(u)}{u}\ \rmd u.
\end{eqnarray*}
Therefore,
$$\int_1^{\csc(\theta)}\frac{\tan^{-1}(yu)}{\sqrt{u^2-1}}\ \rmd u={\rm  Ti}_2(A)-{\rm  Ti}_2(B)+C,$$
where 
\begin{equation}
{\rm  Ti}_2(z)=\int_0^z\frac{\tan^{-1}(u)}{u}\ \rmd u
\label{ti2}
\end{equation}
is the
Arctangent integral function and $C$ is independent of $\theta$.
However, for $\theta=\pi/2$, $A=B$ and the LHS vanishes, so $C=0$.
Now set
$$\theta=\csc^{-1}\sqrt{\frac{b^2-a^2+1}{1-a^2}},\mbox{\hskip .2in}
y=\sqrt{1-a^{-2}}\mbox{\hskip .1in and\hskip
.1in}u=\sqrt{\frac{x^2-a^2+1}{1-a^2}}.$$ 
This gives
\begin{eqnarray}
\int_0^b\frac{\rmd x}{\sqrt{x^2+1-a^2}}\tan^{-1} \left(\frac{\sqrt{x^2+1-a^2}}{a} \right)= \nonumber \\
{\rm  Ti}_2 \left(\frac{\sqrt{b^2+1-a^2}+b}{1+a} \right)-{\rm  Ti}_2 \left(\frac{
\sqrt{b^2+1-a^2}-b}{1+a} \right).
\label{b3}
\end{eqnarray}
We next define
$$
g(a,b)=\int_0^{\infty}\frac{\rmd s}{s^2+1}{\rm Log}[\sqrt{b^2(s^2+1)+1}+a].
$$
Then $g(a,0)=(\pi/2){\rm Log}(1+a)$ and
$$
\frac{\partial}{\partial
b}g(a,b)=b\int_0^{\infty}\frac{[\sqrt{b^2(s^2+1)+1}-a] \ \rmd u}{(b^2s^2+b^2+1-a^2)\sqrt{b^2(s^2+1)+1}} =
\frac{\tan^{-1}(\sqrt{b^2+1-a^2}/a)}{\sqrt{b^2+1-a^2}}.
$$
Reintegration using (\ref{b3}) gives
$$g(a,b)=\frac{\pi}{2}\ln(1+a)+{\rm  Ti}_2\left(\frac{\sqrt{b^2+1-a^2}+b}{1+a} \right)-{\rm  Ti}_2 \left(\frac{\sqrt{b^2+1-a^2}-b}{1+a} \right),$$
which is easily transformed into
\begin{eqnarray*}
\fl
\int_0^{\infty}\frac{{\rm Log}[\sqrt{x^2+c^2}+\alpha]}{x^2+1}\ \rmd x&=&\frac{\pi}{2}{\rm Log}[1+\sqrt{c^2-\alpha^2}] \\
&&+{\rm  Ti}_2\left[\frac{\sqrt{c^2-1}+\alpha}{\sqrt{c^2-\alpha^2}+1}\right]+{\rm  Ti}_2
\left[\frac{\alpha-\sqrt{c^2-1}}{1+\sqrt{c^2- \alpha^2}}\right].
\end{eqnarray*}

\end{document}